\documentclass[aps,prc,twocolumn,superscriptaddress]{revtex4-2}
\usepackage{amssymb}
\usepackage{lipsum} 
\usepackage{slashed}
\usepackage{amsmath}
\usepackage{amsfonts}
\usepackage{xcolor}
\usepackage{placeins}
\usepackage{graphicx}

\newcommand{\be}{\begin{equation}}
\newcommand{\ee}{\end{equation}}

\newcommand{\bra}{\langle}
\newcommand{\ket}{\rangle}

\newcommand{\bea}{\begin{eqnarray}}
\newcommand{\eea}{\end{eqnarray}}

\newcommand{\Acal}{\ensuremath{\mathcal{A}}}

\newcommand{\Rvec}{\ensuremath{\boldsymbol{R}}}

\newcommand{\rvec}{\ensuremath{\boldsymbol{r}}}

\begin{document}
\title{The Relative Abundance of Correlated Spin-zero Nucleon Pairs}

\author{Raz Yankovich}
\affiliation{The Racah Institute of Physics, The Hebrew University, 
Jerusalem 9190401, Israel}
\author{Ehoud Pazy}
\affiliation{Department of Physics,  NRCN, P.O.B. 9001, Beer-Sheva 84190, Israel}
\author{Nir Barnea}
\affiliation{The Racah Institute of Physics, The Hebrew University, 
Jerusalem 9190401, Israel}

\begin{abstract}
We utilize the generalized contact formalism in conjunction with the Woods-Saxon mean-field description of the long-range part of the nuclear wave function to 
assess the relative prevalence of short-range correlation pairs within atomic nuclei. 
We validate our approach by fitting experimental charge density results and electron scattering experiments to a very good agreement.
Applying our model, we calculate the spin-zero short-range correlations contact ratios. Interestingly, for nuclei with $A>50$, we observe a notable dependence on the neutron-to-proton ratio $N/Z$. Specifically, the probability per nucleon to find neutron-neutron pairs increases, while that of proton-proton pairs decreases, whereas the probability of finding neutron-proton pairs remains relatively constant.
To interpret this isospin symmetry breaking effect, we employ a simple 
model based on generalized Levinger constants, linking it to differences in nuclear proton and neutron radii.

\end{abstract}

\maketitle

Atomic nuclei, despite their intricate nature as many-body systems, can often have their collective properties effectively characterized using a mean-field approach like the nuclear shell model. Such mean-field models can successfully describe low-energy observables such as the nuclear radius, or binding energy, but are often deficient when describing observables sensitive to shorter length scales, where the nuclear wave-function is dominated by short-range correlations (SRCs). 
These SRCs predominantly involve nucleon pairs with large relative momenta 
compared to both the typical nuclear Fermi momentum $k_F\approx 250$ MeV/c and their center-of-mass momentum. Their existence and characteristics have been studied experimentally mainly through electron scattering experiments preformed on a wide range of nuclei,  starting from light elements like He and C up to heavier ones such as Pb, see e.g.  \cite{Subedi08, Korover14, HenSci14}.

It has been discovered that neutron-proton ($np$) SRC pairs are nearly 20 times as prevalent as proton-proton ($pp$) or neutron-neutron ($nn$) pairs \cite{Hen2017, Ciofi15,Subedi08}. Furthermore, it has also been experimentally demonstrated that the number of SRC $np$ pairs grows linearly with that of the minority population \cite{Duer19}. This imbalance is regularly related to the properties of the tensor force \cite{Alvioli08}. The scarcity of spin-zero SRCs composed of $pp$, $nn$ or $np$ pairs makes them much less susceptible to experimental probing. Since experimentally it is much easier to detect protons there exists some data on $pp$ SRCs \cite{Shneor07,Duer19}, however little is known regarding $nn$ and spin-zero $np$ SRCs. 
In this study, we focus on these spin-zero SRCs
aiming to provide a theoretical estimate for their relative abundances and isospin dependence.

The theoretical exploration of SRCs and their isospin dependence in light nuclei 
can be conducted using 
\textit{ab-initio} many-body methods such as the variational Monte Carlo (VMC) method \cite{Carlson15}, Coupled Cluster (CC) theory \cite{Hagen14},  
or the No-Core-Shell-Model \cite{NCSM,NCSM09} method. 
However, applying these methods to study SRCs in heavier nuclei is significantly 
more challenging and is presently limited to closed-shell nuclei with up to about $A=40$ nucleons \cite{Lonardoni17}. 

Several methods have been devised allowing theoretical exploration of SRC physics in heavier nuclei. These methods include the low-order correlation operator approximation \cite{Ryckebusch96,Ryckebusch19,Colle14,Colle15}, the Similarity Renormalization Group \cite{Tropiano21,Tropiano24}, and the Generalized Contact Formalism (GCF) 
\cite{Weiss15,Weiss15PRC}, which specifically is applied in this study.
These theoretical frameworks are all based on two main assumptions: \textit{(a)} There is a separation of scales between SRCs 
and the low energy, long-wavelength, characteristics of the atomic nucleus, 
and \textit{(b)} 
The short-range part of the nuclear wave-function describing two nucleons approaching each other is universal in the sense that it is independent of the particular nuclear state.

The GCF has been proven to be a useful framework for the study of SRCs. 
Employing it, one can derive a set of universal asymptotic relations which do not depend on the specific details of the nuclear interaction. One example is the relation between the asymptotic one-body and two-body nuclear momentum distributions \cite{Weiss15PRC}. 
In addition, the GCF has been successfully utilized to obtain relations between the nuclear contacts and the charge density \cite{Weiss19}, to estimate double-$\beta$ decay
\cite{Weiss22}, and for estimating the entanglement entropy for nuclear SRCs \cite{Pazy23}. It also provides a convenient theoretical 
framework for analyzing electron scattering experiments, see e.g. \cite{Schmidt-Nature2020}.

The GCF aims to describe the many-body wave function
in the limit where two nucleons approach each other.
Utilizing the above assumptions as well as the factorization ansatz \cite{Weiss15PRC,Saar23}, the GCF is based on the following asymptotic many-body wave function
\be\label{eq:ansatz}
  \Psi\xrightarrow[r_{ij}\rightarrow 0]{}
             \sum_\alpha\varphi_\alpha(\rvec_{ij})
             \Acal^\alpha\left(\Rvec_{ij},\{\rvec_k\}_{c\not=i,j}\right)~.
\ee
Here, $\varphi_\alpha(\rvec_{ij})$ is a set of universal two-body wave functions 
that depends on $\rvec_{ij}$, the relative distance between the particle pair, and 
the label $\alpha$ that stands for the quantum numbers characterizing the 2-body state.
Specifically, $\alpha=\{(ls)jtt_z\}$ where $l,s$ are the pair's orbital angular momentum, 
and spin, $j$ the total angular momentum, $t$ the isospin, and $t_z$ its $z$ component.
The universal two-body function $\varphi_\alpha$ is multiplied by the particular - state dependent - function $\Acal^\alpha$ which describes the remaining, spectator, $A-2$ nucleons  \cite{Weiss15PRC}. 
$\Acal^\alpha$ depends on the coordinates of the spectator nucleons as well as on the center of mass coordinate of the correlated pair $\Rvec_{ij}$. 

The asymptotic form of the many-body wave function Eq. \eqref{eq:ansatz} implies that the matrix element of any short-range 2-body operator $\hat O_{ij}$ is a product of 
the universal matrix elements $\bra \varphi_\alpha | \hat O_{12} | \varphi_\beta\ket$, and
the ``nuclear contact coefficients'' \cite{Weiss15PRC}
\be\label{eq:contact}
   C^{\alpha \beta}=\frac{A(A-1)}{2}\bra \Acal^\alpha | \Acal^\beta \ket~.
\ee 

The Contact, originally introduced by Tan \cite{Tan08}, is a measure for the probability of 
finding two particles close to each other. More specifically, if the universal states $|\varphi_\alpha\ket$ are properly normalized \cite{Weiss18}, the diagonal nuclear 
contact $C^{\alpha \alpha}$ counts the number of
correlated nucleon pairs with quantum numbers $\alpha$.
Aiming to study the abundances of spin-zero SRC pairs, in the following we will 
mainly focus on the diagonal $s=0$ contacts.

Although the contact plays a major role in the theory of SRCs,
due to scale separation  
it fundamentally represents a long-wavelength property of the nuclear system \cite{Torres21}. 
Consequently, it is anticipated that the values of the contact can be inferred from mean-field, shell-model, calculations. This was indeed demonstrated in \cite{Torres21,Weiss22} as well as in \cite{Tropiano21,Tropiano24}. Building upon these successful examples, 
we utilize here a mean-field model as a low-resolution approximation to describe the $A$-body nuclear wave function $|\Psi_A\ket$. In the long-wave length limit $|\Psi_A\ket$ is represented as a single Slater determinant $|\Phi_A\ket$, with single-particle orbitals computed using the phenomenological Woods-Saxon (WS) potential. Specifically, we adopt the "universal" parameterization of the WS potential, as described in \cite{Dudek82}.

Up to a constant normalization factor, the $A$-body contact $C^{\alpha\beta}(A)$ can be determined by evaluating the expectation value of the short-range operator 
$ \hat{O}_{C}^{\alpha\beta}=\sum_{i<j}{\hat{\delta}(\rvec_{ij})\hat{P}_{\alpha\beta}}$,
with $\hat{P}_{\alpha\beta}=|\alpha\ket\bra\beta|$ a projection operator on the two particle states defined by the quantum numbers $\alpha,\beta$.
Employing the asymptotic many-body wave function Eq. \eqref{eq:ansatz}, the
expectation value reads
\be
  \label{eq:A_body_c}
  \bra \Psi_A | \hat{O}_{C}^{\alpha\beta}| \Psi_A \ket 
  \cong   
  C^{\alpha\beta}(A)
  \bra{\varphi_{\alpha}(\rvec)}|\hat{\delta}(\rvec)\hat{P}_{\alpha\beta}
  |{\varphi_{\beta}(\rvec)}\ket~.
\ee
To overcome the issue of the normalization factor one 
can either fit the contacts to available experimental and computational data
or consider contact ratios, \textit{i.e.}
the ratio of contacts between two nuclei, say $A$ and $B$.
Due to the universality of the two-body SRC functions, for the latter 
case one reads
\be\label{c_ratios}
   \frac{\bra \Psi_A|\hat{O}_{C}^{\alpha\beta}|\Psi_A\ket}
        {\bra\Psi_{B}|\hat{O}_{C}^{\alpha\beta}|\Psi_{B}\ket}
   =
   \frac{C^{\alpha\beta}(A)}{C^{\alpha\beta}(B)}~.
\ee

Eq.(\ref{c_ratios}) implies that the contact ratio is independent of the specific form of the universal wave function describing the correlated pair, and instead relies solely on the quantum numbers that define its spin and isospin. 

Computing the contact  
demands a good approximation for the spectator function.
Given that the spectator function depends on the $A-2$ nucleons not involved in the SRC, employing the mean-field approximation of the many body wave function
$|\Phi_A\ket$ discussed above seems to be justified. 
Specifically we assume that
\be \label{eq:shell}
   \frac{C^{\alpha\beta}(A)}{C^{\alpha\beta}(B)}
   =
   \frac{\bra \Psi_A|\hat{O}_{C}^{\alpha\beta}|\Psi_A\ket}
        {\bra\Psi_{B}|\hat{O}_{C}^{\alpha\beta}|\Psi_{B}\ket}
   \cong
   \frac{\bra \Phi_A|\hat{O}_{C}^{\alpha\beta}|\Phi_A\ket}
        {\bra\Phi_{B}|\hat{O}_{C}^{\alpha\beta}|\Phi_{B}\ket}~.
\ee

In this study, we will exclusively examine the diagonal contacts associated with the leading channels contributing to SRCs, which include the deuteron channel ($j = 1$, $t=0$), and center specifically on the singlet spin-zero $t=1$ $pp$, $np$, and $nn$ s-wave channels ($l = s = j = 0$). 
Therefore, our analysis will focus solely on the four corresponding contacts, which 
for brevity will be denoted as $C_{np}^1(A)$ for the $A$-body deutron channel contact, and 
as $C_{pp}^0(A),C_{np}^0(A),C_{nn}^0(A)$ for the spin-zero contacts.

So far, nuclear contact values have been extracted from various sources: experimental photo-absorption cross-section data \cite{Weiss16}, electron scattering data \cite{Weiss18,Weiss19}, and many-body VMC calculations \cite{Weiss18,Torres21}. 
Here, we employ the mean-field WS potential to calculate these contacts. However, before presenting our main results, it is essential to assess the reliability of our approach. To this end we will compare our resulting contact values with \textit{(a)} contacts obtained from charge density measurements \cite{Weiss19}, and \textit{(b)} inclusive electron scattering cross-section ratios \cite{Fomin12,Fomin17}.

\begin{figure}
\centering
\includegraphics[scale=0.55]{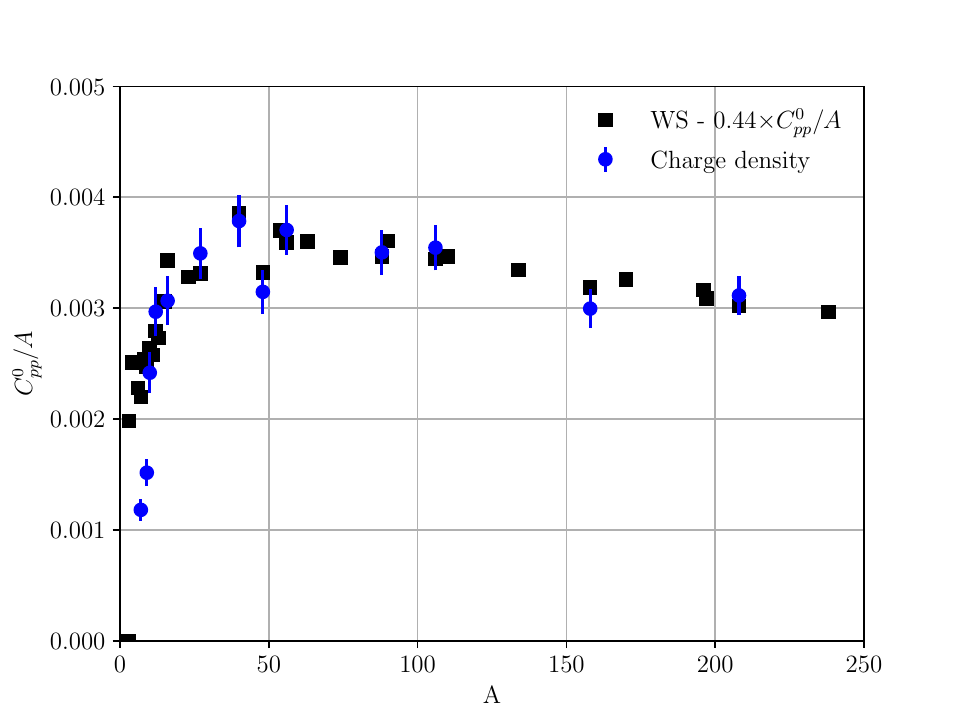}
\caption{The spin-zero contacts $C_{pp}^0(A)$ normalized by $A$. A Comparison between 
contacts calculated with the WS model (black squares) 
and contacts extracted from charge density data \cite{Weiss19} (blue circles).}
\label{fig:cpp}
\end{figure}

In Fig.(\ref{fig:cpp}) we compare the spin-zero contacts $C_{pp}^0(A)$ derived from the WS mean-field 
calculations with  
contacts extracted from experimental charge density measurements \cite{Weiss19}.
To fix the contacts we employed an overall normalization factor $\eta=0.44$, which gives an excellent agreement between our model and the charge density data.
From the figure it is evident that, as expected, in all but the lightest nuclei the contact is more or less linear with the number of nucleons.  
It is also interesting to note that the model manages to reproduce not only the general trends of the data, that is fast growth for small $A$ and slow decline for the heavier nuclei, it is even  able to reproduce the detailed structure appearing around $A\approx 50$.

\begin{figure}
\centering
\includegraphics[scale=0.5]{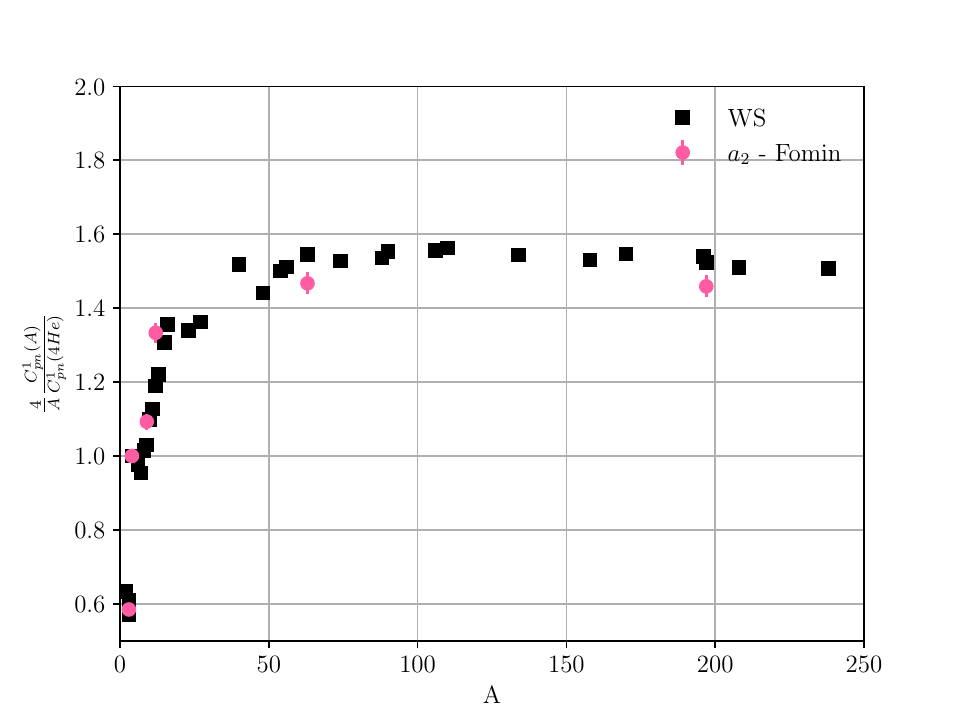}
\caption{A comparison between $a_2$ ratios extracted from inclusive electron scattering results \cite{Fomin17} (pink circles), and contact ratios, Eq.  \eqref{a2_ratio}, calculated 
with the WS model (black squares).}
\label{fig:a2}
\end{figure}

Electron scattering experiments performed at the kinematic region of
$1.4 \leq x_B < 2$ are dominated by 2-body nuclear SRCs \cite{Fomin17}. In fact, in this region
the inclusive $(e,e')$ cross-section is proportional to the deutron cross-section
$\sigma_2(e,e')$, and the
cross-section ratio $a_2\equiv (2/A)(\sigma_A(e,e')/\sigma_2(e,e'))$ is roughly a constant.
In view of the deutron channel dominance, and following the prescriptions presented in Refs. \cite{Weiss18} and \cite{Ryckebusch19}, we estimate the $a_2$ ratio between two nuclei $A$, and $B$, through the contact ratio
\be\label{a2_ratio}
  \frac{a_2(A)}{a_2(B)} \cong \frac{B}{A}\frac{C^1_{np}(A)}{C^1_{np}(B)}~.
\ee 
Here $A,B$ also stand for the number of nucleons in each nucleus.
In Fig. (\ref{fig:a2}) we present the calculated $a_2$ ratios for several stable nuclei using $^4$He as a reference.
Also shown are the experimental results of \cite{Fomin17}.
The comparison between our model and the experimental results indicates
that it is a fairly accurate model.

\begin{figure}
\centering
\includegraphics[scale=0.5]{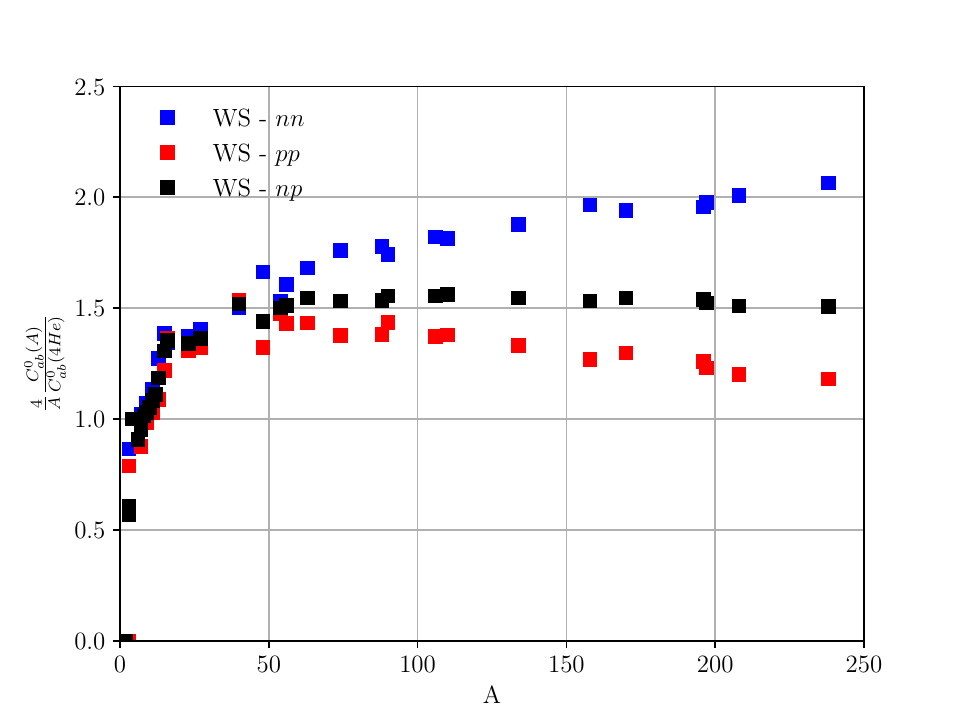}
\caption{The nuclear spin-zero contacts $C^0_{nn}$ (blue), $C^0_{pp}$ (red), and $C^0_{np}$ (black),
calculated with the WS model. 
The contacts are normalized by $A$, and presented relative to the corresponding $^4$He contacts.
}
\label{fig:SpinZeroContacts}
\end{figure}

The relative simplicity of our mean-field model makes it a convenient tool for
a systematic study of nuclear SRCs. Specifically, it allow us to explore
the evolution of the three spin-zero nuclear contacts along the table of nuclei,
and study the effect of neutron $N$ to proton $Z$ imbalance on these contacts.
In Fig. \ref{fig:SpinZeroContacts} we present these contacts as a function of $A$.
Due to isospin symmetry, we anticipate $C_{pp}^0(A)=C_{np}^0(A)=C_{nn}^0(A)$ for symmetric $N=Z$ isospin zero nuclei. This symmetry holds true notably for light nuclei with $A<50$. However, as $A$ approaches approximately 50, the behavior of these contacts starts to change. Specifically, the three spin zero contacts begin to diverge from each other with increasing $A$. The $nn$ contact per nucleon 
continues to grow, the $np$ contact which in our model is proportional to the deutron contact presented in Fig. \ref{fig:a2}
is roughly a constant, and the $pp$ contact shows a slight decline.

\begin{figure}
\centering
\includegraphics[scale=0.5]{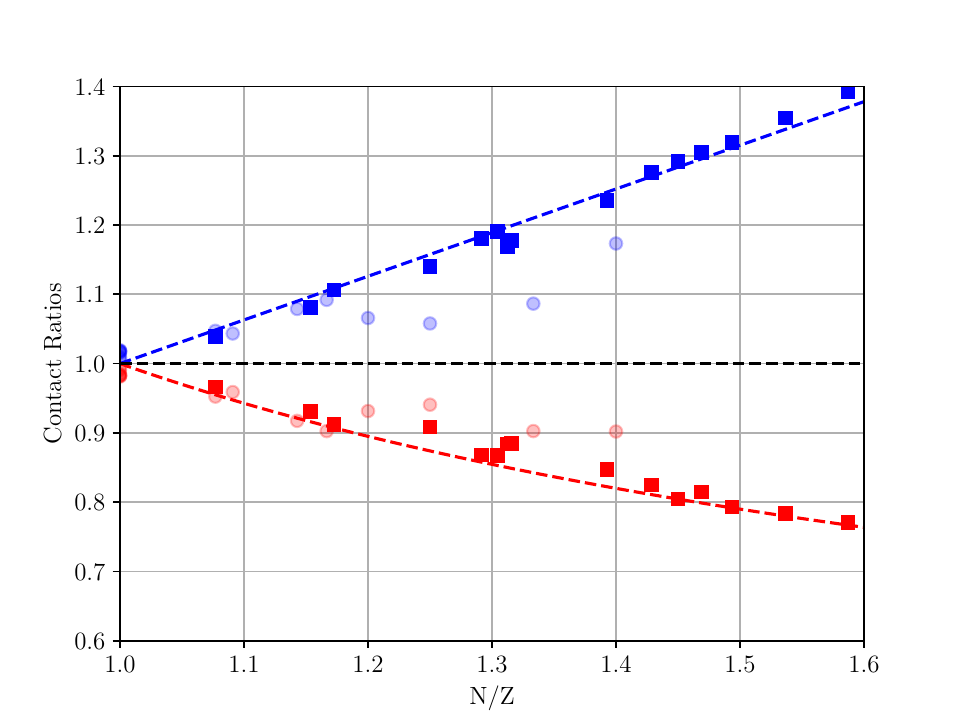}
\caption{The nuclear spin-zero contact ratios $C^0_{pp}/C^0_{np}$ (red), and $C^0_{nn}/C^0_{np}$ (blue) as a function of the neutron excess $N/Z$.
Nuclei with $A\ge 50$ are represented by squares, while nuclei with $A<50$ 
are represented by faded circles.
The dashed lines are the fits presented in \eqref{eq:fit}.
The contacts were calculated using the WS model. }
\label{fig:Contact_ratios}
\end{figure}

To further examine the impact of population imbalance on the nuclear spin-zero contacts
we present in Fig. \ref{fig:Contact_ratios} the ratios 
$C^0_{pp}/C^0_{np}$, and $C^0_{nn}/C^0_{np}$ as a function of the neutron excess $N/Z$.
As expected, for $N=Z$ the nuclear contacts are equal due to isospin symmetry. 

Interestingly, for nuclei with $A>50$ the contact ratio $ C^0_{nn}(A)/C^0_{np}(A)$ 
presents a rather linear dependence on $N/Z$, and 
same for $C^0_{pp}(A)/C^0_{np}(A)$ with $Z/N$.
A linear fit for these ratios yield
\begin{align}\label{eq:fit}
    \frac{C^0_{nn}}{C^0_{np}} &\approx \left[ 1 + \beta \left(\frac{N}{Z}-1\right)\right]
    \cr
    \frac{C^0_{pp}}{C^0_{np}} &\approx \left[ 1 + \beta \left(\frac{Z}{N}-1\right)\right]~,
\end{align}
with $\beta \approx 0.63$. 
Surprisingly, these expressions are symmetric under the exchange of neutrons
with protons (and of course $N\leftrightarrow Z$).

The microscopic origin of this phenomenon can be attributed either to the increasing dominance of the Coulomb force in heavier nuclei or to the significant dependence of the WS potential on the neutron-proton imbalance $(N-Z)$.

When comparing the spin-zero contacts of the two $Z=20$ isotopes, $^{40}$Ca and $^{48}$Ca 
(shown just below $A=50$ in Fig. \ref{fig:SpinZeroContacts}), it becomes evident that it is the neutron excess terms in the WS potential, rather than the Coulomb force, that break the isospin symmetry of these contacts.

To gain better insight and a more intuitive understanding of the results regarding isospin symmetry breaking, we introduce a simplistic model for estimating the contacts. This approach follows some of the arguments leading to the derivation of the quasi-deuteron model proposed by Levinger \cite{Levinger51}.

Levinger studied the nuclear photo-absorption cross section $\sigma_A(\gamma)$ in the photon energy range of a few hundred MeV. He suggested that the dipole reaction mechanism could be attributed to deuteron-like nucleon pairs within the nucleus, leading to a cross section proportional to the deuteron photo-absorption cross section $\sigma_2(\gamma)$. By counting these pairs, he derived the 
relation $\sigma_A(\gamma)=L (NZ/A)\sigma_2(\gamma)$ between the two, with 
$L$ being the Levinger constant.

Subsequently, Weiss et al. \cite{Weiss15} demonstrated that the Levinger constant can be linked to the nuclear deutron channel contact $C_{np}^1$. Extending this idea we introduce the constant $L^{s}_{ab}$ which is a generalized Levinger constant, independent of the specific nucleus. Based on Levinger's  reasoning, considering nucleon pairs of  total spin $s$ composed of nucleons $a$ and $b$ (where $a,b$ can be either $n$ or $p$), the number of short-range correlated pairs must take the form:
\be\label{eq:GenerlaizedL}
    C^{s}_{ab}(A) = L^{s}_{ab} ~\rho_{a}\rho_{b}\Omega_{ab}~,
\ee
where $\rho_{a},\rho_b$ denote the number densities of nucleons of type $a$ and $b$ respectively,
and $\Omega_{ab}$ represents the relevant nuclear volume. 

This model provides a framework for estimating the contacts $C^{s}_{ab}(A)$ using 
the nuclear densities and volumes.

Let $R_p$ denote the root mean square (rms) proton radius. Omitting unnecessary constants, we estimate the proton volume as $\Omega_p\approx R_p^3$, leading to a proton density $\rho_p\approx Z/R_p^3$. 
Similarly, for neutrons, let $R_n$ be the rms neutron radius, giving $\Omega_n\approx R_n^3$
and $\rho_n\approx N/R_n^3$.

For the common $np$ volume, we use the rms matter radius $R$, approximating $\Omega_{np}\approx R^3$.
Consequently, the spin-zero $nn,np$, and $pp$ contacts are expressed as:
\begin{align}
    C_{nn}^0(A) &= L^{0}_{nn} \frac{N^2}{R_N^3}
    \cr
    C_{np}^0(A) &= L^0_{np} \frac{N Z}{R_N^3 R_P^3}{R^3}
    \cr
    C_{pp}^0(A) &= L^0_{pp} \frac{Z^2}{R_P^3}~.
\label{eq:Levinger}
\end{align}

\begin{figure}
\centering
\includegraphics[scale=0.5]{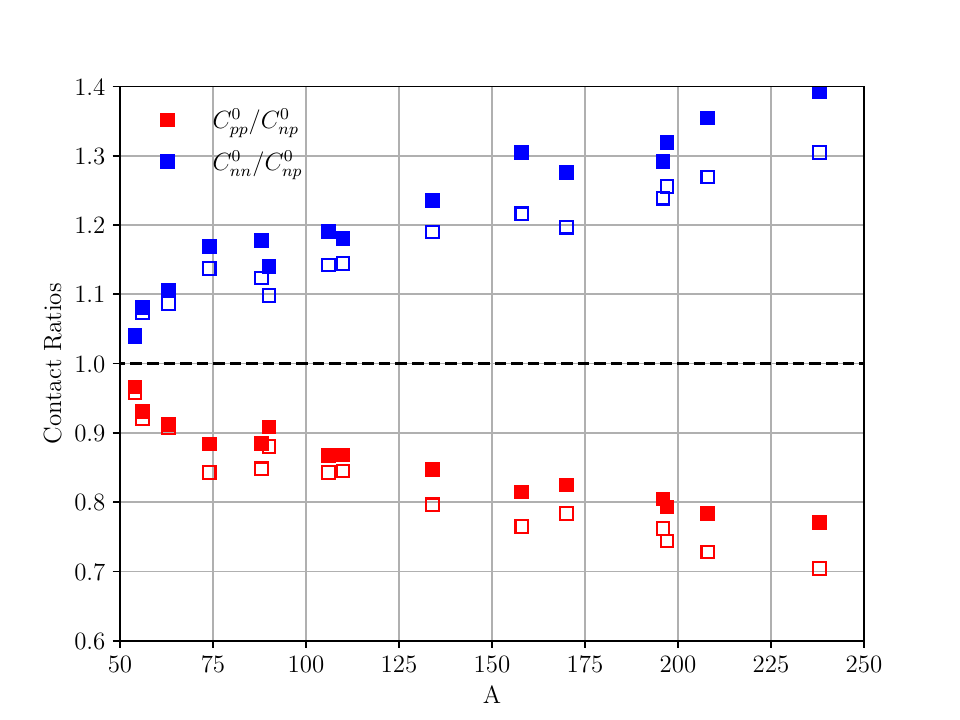}
\caption{The nuclear spin-zero contact ratios $C^0_{pp}/C^0_{np}$ (red), and $C^0_{nn}/C^0_{np}$ (blue) as a function of $A$.
Comparison between contact ratios calculated with the WS mean-field model (full squares), and the generalized Levinger model (empty squares).}
\label{fig:compare_contact_ratio}
\end{figure}

Considering now a symmetric $N=Z$ isospin zero nucleus, due to isospin symmetry
one must conclude that the
generalized spin-zero Levinger constants must all be equal, i.e. 
${L^0_{nn}}={L^0_{np}}=L^0_{pp}$. At first sight this conclusion seems a bit odd, as
one would naively expect that the number of $np$ pairs is roughly twice the number
of $nn$ or $pp$ pairs. However, this factor of 2 disappears
when taking into account that the $np$ pair must be in an $s=0$ state.
Based on this argument and using the above expressions Eqs. (\ref{eq:Levinger}), 
the spin-zero contact ratios take the form
\begin{align}\label{eq:Levinger_conatct_ratios}
    \frac{C_{nn}^0(A)}{C_{np}^0(A)} 
    &\approx 
     \frac{N}{Z} {\left(\frac{R_P}{R}\right )}^3
    \cr
    \frac{C_{pp}^0(A)}{C_{np}^0(A)} 
    &\approx 
    \frac{Z}{N} {\left(\frac{R_N}{R}\right )}^3~.
\end{align}
We use the WS potential to calculate the rms neutron, proton, and matter radii $R_N, R_P, R$. Inserting the obtained values into Eq. (\ref{eq:Levinger_conatct_ratios})
we compare in Fig. \ref{fig:compare_contact_ratio} these simplistic contact ratios,
which emerge from the generalized Levinger model to our WS contact calculations.
It can be seen that the, parameter free, generalized Levinger model reproduces 
in a fairly accurate manner the calculated results.

In summary, using a mean-field, shell-model, description of the nuclear wave-function with the Woods-Saxon potential, we computed the SRC contacts and successfully fitted them to available data. 
Specifically, for the spin-zero contacts, we found a pronounced isospin symmetry breaking effect, which we attribute to the varying radii of protons and neutrons, and is explained through a generalized Levinger model. A significant outcome of this research is the acknowledgment of the sensitivity of these contacts to difference in nuclear radii among distinct nuclei.

Considering the rapid advancements in experimental studies of SRCs, we expect that validating these findings will be feasible in the near future.

\section*{Acknowledgement}
We wish to thank R. Weiss for careful reading of the manuscript and for his valuable comments. 
This work was supported by the Pazy Foundation 
and by the Israel Science Foundation, grant number 1086/21.



\begin{thebibliography}{99}
\bibitem{Subedi08}
R. Subedi et al., Science {\bf 320}, 1476 (2008).

\bibitem{Korover14}
I. Korover, et al., Phys. Rev. Lett. {\bf 113}, 022501 (2014).

\bibitem{HenSci14}
O. Hen, et al., Science {\bf 346}, 614 (2014). 

\bibitem{Hen2017}
O. Hen, G. A. Miller, E. Piasetzky, and L. B. Weinstein,
Rev. Mod. Phys. {\bf 89}, 045002 (2017).

\bibitem{Ciofi15}
C. Ciofi degli Atti, Phys. Rep. {\bf 590}, 1 (2015)

\bibitem{Duer19}
M. Duer, et. al., Phys. Rev. Lett. {\bf 122}, 172502 (2019).

\bibitem{Alvioli08}
M. Alvioli, C. Ciofi degli Atti, and H. Morita, Phys. Rev. Lett. {\bf 100}, 162503

\bibitem{Shneor07}
R. Shneor, et. al. Phys. Rev. Lett. {\bf 99}, 072501 (2007). 

\bibitem{Carlson15}
J. Carlson, S. Gandolfi, F. Pederiva, S.C. Pieper, R. Schiavilla, K.E. Schmidt, R.B. Wiringa, Rev. Mod. Phys. {\bf 87}, 1067 (2015).

\bibitem{Hagen14}
G. Hagen, T. Papenbrock, M. Hjorth-Jensen, and D J Dean,
Rep. Prog. Phys. {\bf 77}, 096302 (2014).

\bibitem{NCSM}
P. Navr{\'a}til, J. P. Vary, and B. R. Barrett,
Phys. Rev. C {\bf 62}, 054311 (2000).

\bibitem{NCSM09}
P. Navr{\'a}til, S. Quaglioni, I. Stetcu, and B. R. Barrett,
{J. Phys. G} {\bf 36}, 083101 (2009).

\bibitem{Lonardoni17}
D. Lonardoni, A. Lovato, S.C. Pieper, R.B. Wiringa, Phys. Rev. C {\bf 96}, 024326 (2017).  


\bibitem{Ryckebusch96}
J. Ryckebusch, Phys.Lett. B {\bf 383}, 1 (1996).

\bibitem{Ryckebusch19}
J. Ryckebusch, W.Cosyn, S. Stevens, C. Casert and J. Nys, Phys. Rev. C {\bf 100}, 054620 (2019).

\bibitem{Colle14}
C. Colle, W. Cosyn, J. Ryckebusch, and M. Vanhalst, Phys. Rev. C {\bf 89}, 024603 (2014).

\bibitem{Colle15}
 C. Colle, O. Hen, W. Cosyn, I. Korover, E. Piasetzky, J. Ryckebusch, and L.B. Weinstein,
hys. Rev. C {\bf 92}, 024604 (2015).

\bibitem{Tropiano21}
A. J. Tropiano, S. K. Bogner, and R. J. Furnstahl,
Phys. Rev. C {\bf 104}, 034311 (2021).

\bibitem{Tropiano24}
    A.J. Tropiano, S.K. Bogner, R.J. Furnstahl, M.A. Hisham,
    A. Lovato, and R. B. Wiringa,
    Phys. Lett. B {\bf 852}, 138591 (2024).

\bibitem{Weiss15PRC}
R. Weiss, B. Bazak, and N. Barnea, Phys. Rev. C {\bf 92}, 054311 (2015).

\bibitem{Weiss15}
R. Weiss, B. Bazak, and N. Barnea, Phys. Rev. Lett. {\bf 114}, 012501 (2015).

\bibitem{Weiss19}
R. Weiss, A. Schmidt, G. A. Miller, N. Barnea,  Phys. Lett. B {\bf 790}, 484 (2019).

\bibitem{Weiss22}
R. Weiss, P. Soriano, A. Lovato, J. Menendez, and R. B. Wiringa,
Phys. Rev. C {\bf 106}, 065501 (2022).

\bibitem{Pazy23}
E. Pazy, Phys. Rev. C {\bf 107}, 054308 (2023).

\bibitem{Schmidt-Nature2020}
A. Schmidt, et. al. , 
  Nature {\bf 578}, 540 (2020).

\bibitem{Saar23}
S. Beck, R. Weiss, and N. Barnea, 
Phys. Rev. C {\bf 107}, 064306 (2023).

\bibitem{Tan08}
S. Tan, Ann. Phys. (N.Y.) {\bf  323}, 2952 (2008); Ann. Phys.
(N.Y.) {\bf 323}, 2971 (2008); Ann. Phys. (N.Y.) {\bf 323}, 2987
(2008).   

\bibitem{Weiss18}
R. Weiss, R. Cruz-Torres, N. Barnea, E. Piasetzky and O. Hen, Phys. Lett. B {\bf 780}, 211 (2018).

\bibitem{Torres21}  
R. Cruz-Torres et al., Nature Physics {\bf 17}, 306 (2021).

\bibitem{Dudek82}
J. Dudek, Z. Szymanski, T. R. Werner, A. Faessler, and C. Lima,
Phys. Rev. C {\bf 26}, 1712 (1982).

\bibitem{Weiss16}
R. Weiss, B. Bazak, and N. Barnea, Eurp. Phys. J. A, {\bf 52}, 92 (2016). 
\bibitem{Fomin17}
N. Fomin, D. Higinbotham,M. Sargsian, and P. Solvigno,  Annu. Rev. Nucl. Part. Sci. {\bf 67},129, (2017). 

\bibitem{Fomin12}
N. Fomin et al., Phys. Rev. Lett. {\bf  108}, 092502 (2012).

\bibitem{Levinger51}
J. S. Levinger, Phys. Rev. {\bf 84}, 43 (1951).

\bibitem{Alvioli13}
M. Alvioli, C. Ciofi degli Atti, L. P. Kaptari, C. B. Mezzetti, and H. Morita
Phys. Rev. C {\bf 87}, 034603 (2013).







%
%
%
%
%
%



\end{thebibliography}
\end{document}